%% file: Main.tex
\documentclass[%
 reprint,
 amsmath,amssymb,
 aps,
]{revtex4-2}

\usepackage{CJK} 
\usepackage{amsmath}
\usepackage{here}
\usepackage{mathptmx} 
\usepackage[dvipdfmx]{graphicx}
\usepackage{bm} 
\usepackage{color}
\usepackage{lipsum}

\date{February 11, 2021}
\usepackage{newtxtext,newtxmath}

\newcommand{\argmin}{\mathop{\rm arg~min}\limits}

\usepackage{dcolumn}
\usepackage{bm}

\begin{document}

\preprint{APS/123-QED}

\title{Fast acquisition of spin-wave dispersion by compressed sensing}
\author{Ryo Kainuma$^{1}$}
\author{Keita Matsumoto$^{1, 2}$}
\author{Takuya Satoh$^{1}$}
\affiliation{%
 $^{1}$Department of Physics, Tokyo Institute of Technology, Tokyo 152-8551, Japan \\
 $^{2}$Department of Physics, Kyushu University, Fukuoka 819-0385, Japan
}

\begin{abstract}
    For the realization of magnonic devices, spin-wave dispersions 
    need to be identified. Recently, the time-resolved pump-probe imaging 
    method combined with the Fourier transform was demonstrated for obtaining 
    the dispersions in the lower-wavenumber regime. However, the measurement takes 
    a long time when the sampling rate is sufficiently high. Here, we demonstrated 
    the fast acquisition of spin-wave dispersions by using the compressed sensing 
    technique. Further, we quantitatively evaluated the consistency of the results. 
    Our results can be applied to other various pump-probe measurements, 
    such as observations based on the electro-optical effects.
\end{abstract}

\maketitle

\input{intro.tex}

\input{method_experiment.tex}

\input{method_CS.tex}

\input{result.tex}

\input{conclusion.tex}

\section*{Acknowledgments}
We would like to thank K. Hukushima and T. Ishikawa for valuable suggestions. 
This study was supported by the Japan Society for the Promotion of Science (JSPS)
KAKENHI (Grants No. JP19H01828, No. JP19H05618, No. JP19J21797, No. JP19K21854,
and No. JP26103004). K. M. would like to thank the Research 
Fellowship for Young Scientists by the JSPS.

\bibliography{CS}

\end{document}

%% file: intro.tex
Spin  waves are collective modes of spin precession in magnetically ordered materials. 
They are considered promising
information carriers in the field of magnonics, 
because they can propagate over a long distance without Joule heating \cite{gurevich1996magnetization, stancil2009spin, kruglyak2010magnonics,chumak2015magnon}. 
Various devices such as spin wave switches \cite{lenk2011f}, magnonic-logic circuits \cite{schneider2008realization, kanazawa2017role}, spin wave-assisted recorders, \cite{lenk2011f,seki2013spin},
and low-magnetic-field sensors \cite{lee2010nanoscale} require the spatial control of spin waves.

The propagation characteristics of spin waves are manifested in
their dispersion relation.
The higher-wavenumber regime is governed by exchange interactions.
In contrast, lower-wavenumber spin waves are governed 
by magnetic dipole interactions and are called magnetostatic waves \cite{damon1961magnetostatic,damon1965propagation,hurben1995theory,hurben1996theory}.
Since the magnetostatic waves are suitable for long-distance propagation, 
further investigation of dispersion relations in the lower-wavenumber regime is indispensable \cite{stancil2009spin}.

Experimental techniques for acquiring dispersion relations of spin waves are being actively studied.
For example, inelastic neutron scattering \cite{mook1985neutron,liu1996exchange}
and spin-polarized electron energy loss spectroscopy \cite{vollmer2003spin} have been demonstrated.
However, these methods are suitable for observing the higher-wavenumber region, 
rather than observing the lower-wavenumber region of the dispersion.

Recently, a method called spin-wave tomography (SWaT), which used time-resolved pump-probe measurements
and the Fourier transform to visualize the dispersion relations of 
spin waves in the lower-wavenumber region was demonstrated \cite{hashimoto2017all,hashimoto2018phase}.
Further, similar measurements in metals were performed using the magneto-optical Kerr effect \cite{kamimaki2017reciprocal}.
In the SWaT method, a pump pulse is used to impulsively excite the spin wave, and a probe pulse with 
a time delay is used to detect the change in magnetization.
By using an ultrashort pulsed laser as a pump pulse, spin waves in a wide frequency range can be excited simultaneously.
Moreover, by focusing the pulses, spin waves in a wide wavenumber range can be excited.
Wavenumber-resolved measurements can be made by spatially scanning a sample with the focused probe pulses \cite{kamimaki2017reciprocal} 
or imaging a large area without focusing the probe pulses \cite{hashimoto2017all,hashimoto2018phase}.
Therefore, this method is useful for observing the dispersions over a wide region in the wavenumber ($k$)--frequency ($f$) space.

For observing the dispersion, a sufficiently high sampling rate must be maintained for the time-resolved measurement.
This is because of the risk of folding noise due to the nature of the discrete Fourier transform (DFT).
As a result, the measurement time can range from 10 hours to several days.
To search for novel photo-induced dynamics in innumerable materials, the measurement time must be reduced.

In recent years, in experiments such as 
the terahertz imaging \cite{chan2008single},
the NMR spectroscopy \cite{kazimierczuk2011accelerated}, and
the scanning tunneling microscopy and spectroscopy \cite{nakanishi2016compressed}, 
it has been shown that a method called compressed sensing 
can reduce the measurement time. 
Compressed sensing is a signal processing technique
that allows the estimation of a signal from a small amount of data \cite{candes2006robust}. 
The signal estimation can be achieved by 
the least absolute shrinkage and selection operator (LASSO) method, 
a commonly used form of sparse regression \cite{tibshirani1996regression}.

In this letter, we demonstrate the fast acquisition of
the spin-wave dispersions by time-resolved pump-probe 
measurements using compressed sensing. 
Moreover, we quantitatively evaluated the effect of reducing the number of 
measurements in compressed sensing on the results of observations.

%% file: method_experiment.tex
\begin{figure}
\centering
\includegraphics[width=\linewidth]{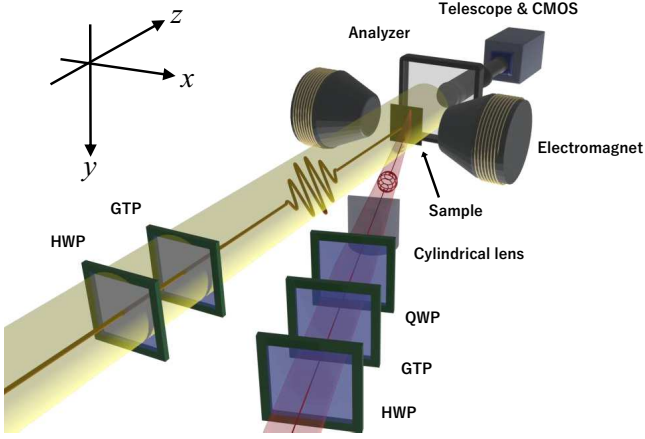}
\caption{Setup of our pump-probe experiment. The circularly polarized pump pulse was focused along a line by a cylindrical lens.
The linearly polarized probe pulse was irradiated on the entire sample without focusing.
HWP: half-wave plate, GTP: Gran Taylor prism, and QWP: quarter-wave plate.}
\label{fig:setup}
\end{figure}

Our experimental setup is shown in Fig. \ref{fig:setup}.
Our sample was a single crystal of (111)-oriented 150 $\mu \mathrm{m}$ thick bismuth-doped rare earth iron garnet
(Gd$_{3/2}$Yb$_{1/2}$BiFe$_5$O$_{12}$) 
grown on a gadolinium gallium garnet substrate by the liquid-phase epitaxy method.
This magnetic material has been widely used for 
investigating laser-induced spin dynamics due to its strong magneto-optical coupling 
\cite{satoh2012directional,parchenko2013wide,yoshimine2014phase,yoshimine2017unidirectional,chekhov2018surface,matsumoto2018optical,matsumoto2020observation}.
The magnetic field of $H_\mathrm{ext}=2440\ \mathrm{Oe}$ was applied in the $x$-direction.

The light pulse for this pump-probe measurement was generated by a Ti:sapphire regenerative amplifier with a pulse duration of $70\ \mathrm{fs}$ and a repetition rate of $1\ \mathrm{kHz}$.
A circularly polarized pump pulse with a central wavelength of $1300 \ \mathrm{nm}$ was focused along a line parallel to the $y$-axis with a cylindrical lens with a fluence of $80 \ \mathrm{mJ\ cm^{-2}}$.
This pump pulse produced an effective magnetic field in the $z$-direction via the inverse Faraday effect \cite{kimel2005ultrafast}, 
and the magnetization saturated in the $x$-direction 
tilted in the $y$ direction.
Since the effective magnetic field was instantaneous, the magnetization then began to precess in the $y$-$z$ plane.
The spin precession excited along the line-shaped pumping spots propagated perpendicular to the line via magnetic dipole interactions.
A linearly polarized pulse with a central wavelength of 800 $\ \mathrm{nm}$ was used 
to probe the $z$-component of magnetization $m_z (\mathbf{r}, t)$ via the Faraday effect.
The Faraday rotation angle was determined from the angle of the analyzer,
which minimized the intensity of the transmitted probe pulse detected by a 
complementary metal--oxide semiconductor (CMOS) camera.
The time delay between the pump and probe pulses was achieved by a variable optical path difference using a delay stage.
This was changed in increments of $\Delta t = 0.01 \ \mathrm{ns}$.
The maximum value of the delay was set to $T = 3.60 \ \mathrm{ns}$.
Then, we obtained the spatiotemporal waveform of the spin wave.
The waveform was integrated along the $y$-direction to create a one-dimensional waveform $m_z(x, t)$.
The power spectrum that depicts the dispersion curve
in the $k$--$f$ space was obtained 
by the two-dimensional DFT of the spatiotemporal waveform.
Further, micromagnetic simulations were performed to confirm our experimental results (See supplementary data).

%% file: method_CS.tex
Let $N = T/ \Delta t$ be the number of samples in the time domain of the time-resolved 
pump-probe measurement.
Owing to the nature of the DFT, 
the frequency resolution of the spectrum is $\Delta f=1/T$.
In addition, according to Nyquist's theorem, the frequency components above 
$1/(2 \Delta t)$ cannot be observed, and they appear as folding noise.
Therefore, with regard to reducing the measurement time, decreasing $T$ leads to a poor resolution,
and increasing $\Delta t$ increases the risk of the appearance of folding noise.

Compressed sensing is a method of reducing $N$ by taking $\Delta t$ randomly, 
without changing $T$, and estimating the spectrum from such data.
Since $\Delta t$ is not a constant, the DFT does not work well, 
resulting in spectral leakage.
Instead, the spectrum can be estimated by the LASSO method.

In the LASSO method, the spectrum estimation is treated as an inverse problem.
Let $\mathbf{y} = (y(t_1), y(t_2), \cdots , y(t_N))^{\mathrm{T}}$ 
be a waveform sampled with discrete time $t_i\ (i = 1, 2, \cdots, N)$
and $\mathbf{x} = (a(\omega_1), a(\omega_2), \cdots ,a(\omega_M), b(\omega_1), b(\omega_2), \cdots ,b(\omega_M))^{\mathrm{T}}$
are the cosine and sine components of the spectrum for discrete frequencies $\omega_j\ (j=1, 2, \cdots, M)$.
The spectrum to be estimated is the solution $\hat{\mathbf{x}}$ of the following minimization problem
\begin{align}
    \hat{\mathbf{x}} = \argmin_{\mathbf{x}} \left\{ \frac{1}{2N} || \mathbf{y} - A \mathbf{x} ||_2^2 + \lambda || \mathbf{x} ||_1 \right\}. 
    \label{eq:LASSO}
\end{align}
\textcolor{black}{Arg min$\{\cdot \}$
denote the argument of the minimum, an element that minimizes the value in the brackets.
$|| \cdot ||_p $ is a term called the $l_p$ norm of the vector and is defined as $|| \mathbf{x} ||_p = \left(\sum_i x_i^p \right)^{1/p}$.}
The first term on the right-hand side corresponds to the method of least squares.
$A$ is a $N \times 2M$ matrix, which corresponds to 
the inverse Fourier transform: 
\begin{align}
A_{ij} = \begin{cases}
\cos (\omega_{j} t_{i}) \qquad &(j = 1, 2, \cdots , M) \\
\sin (\omega_{j-M} t_{i}) \qquad &(j = M+1,M+2, \cdots, 2M).
\end{cases}
\label{eq:InverseFourier} 
\end{align}
The second term on the right-hand side in Eq. (\ref{eq:LASSO}) imposes a sparsity constraint on 
the solution $\hat{\mathbf{x}}$, and $\lambda$ is a parameter that adjusts the sparsity.
We determined $\lambda$ via five-fold cross-validation \cite{james2013introduction}.
\textcolor{black}{In this method, the elements of $\mathbf{y}$ were randomly divided into five data sets. 
Four of these sets were used to obtain $\hat{\mathbf{x}}$, 
and the other set was used to evaluate the waveform reproduced from $\hat{\mathbf{x}}$.}

Generally, the output of the LASSO is not strictly unique, 
and it depends on the random sampling \cite{haury2011influence}.
Cosine similarities (CS) 
were used to evaluate the consistency of results from different dataset.
The CS of two dispersions is given by
\begin{align}
    \mathrm{CS} = \frac{\mathbf{f}\cdot \mathbf{g}}{|\mathbf{f}||\mathbf{g}|}
\end{align}
where $\mathbf{f}$ and $\mathbf{g}$ are vectorized data of the dispersion relations.
\textcolor{black}{Here, $\mathbf{f}$ is the dispersion calculated by LASSO from the $N=361$ data
as the most ideal dispersion available from the present data
and $\mathbf{g}$ is the dispersion with reduced $N$ to be compared.}

For the DFT method, $N$ was reduced by taking 
$\Delta t$ as $0.01\ \mathrm{ns}$ multiplied by the divisors of 360
without changing $T$.
For the LASSO method, random sampling with $N=10,\ 20,\ \cdots ,\ 360$ was performed in ten ways each, 
and the mean and standard deviation of the CSs were calculated.
Moreover, the points $t = 0,\ 0.01,\ 3.60\ \mathrm{ns}$ were
always sampled to maintain the frequency resolution and the Nyquist frequency. 

%% file: result.tex
\begin{figure*}
    \centering
    \includegraphics[width=\linewidth]{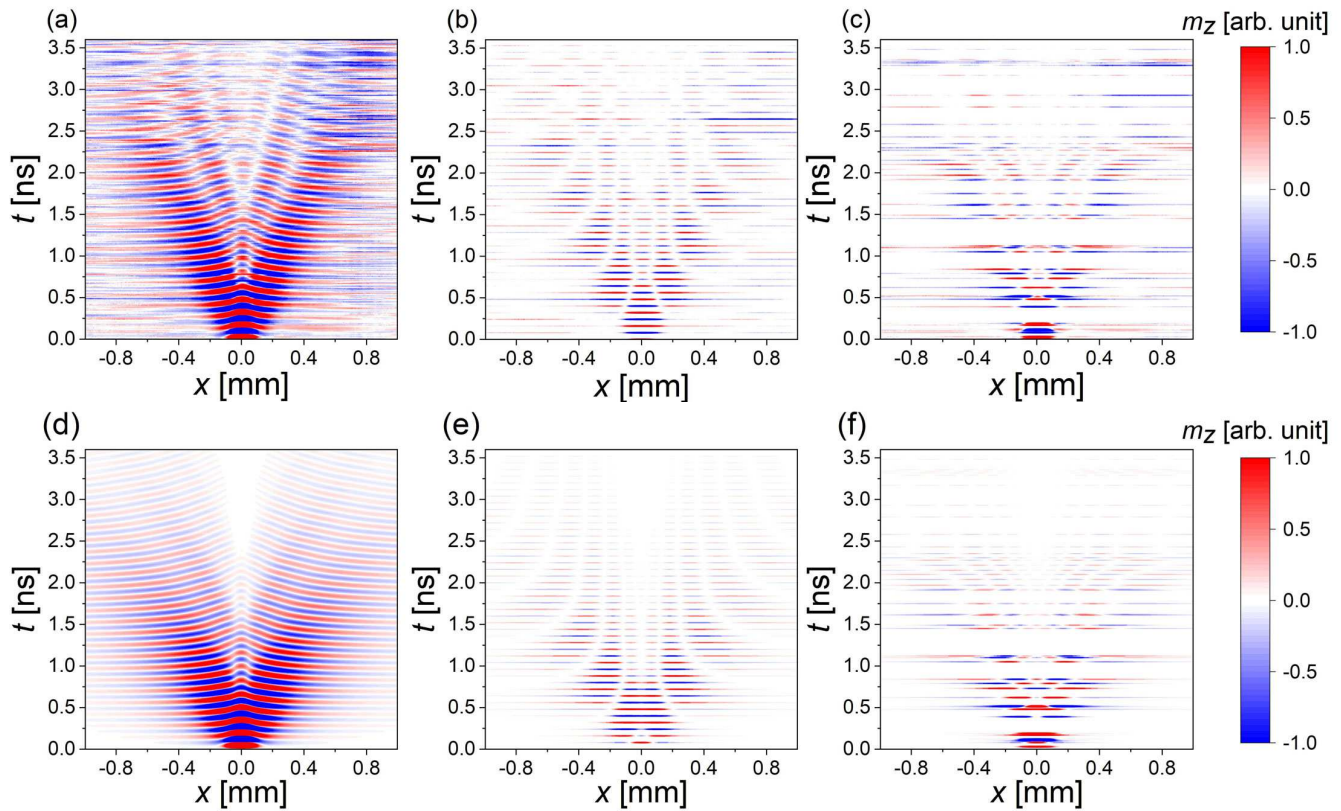}
    \caption{Spatiotemporal waveforms of spin wave observed by time-resolved pump-probe magneto-optical imaging method. (a) the whole data with $N=361$. (b) waveform sampled from (a)
    with $N=46$ by taking $\Delta t = 0.08 \ \mathrm{ns}$. (c) waveform sampled from (a) at random in time domain
    with $N=46$. (d)--(f) waveforms simulated via MuMax3 corresponding to (a)--(c), respectively.}
    \label{fig:STmap}
\end{figure*}

Figure \ref{fig:STmap}(a) shows the entire spatiotemporal waveform
of the spin wave that we observed in our experiments with $N=361$
at increments of $\Delta t = 0.01 \ \mathrm{ns}$. 
Figure \ref{fig:STmap}(b) shows a waveform dataset with $N=46$ by taking $\Delta t = 0.08 \ \mathrm{ns}$,
and Fig. \ref{fig:STmap}(c) represents a dataset with $N=46$ by taking $\Delta t$ at random.
Further, Figs. \ref{fig:STmap}(d)--(f) shows the simulated waveforms
correspond to Figs. \ref{fig:STmap}(a)--(c).

\begin{figure*}
    \centering
    \includegraphics[width=\linewidth]{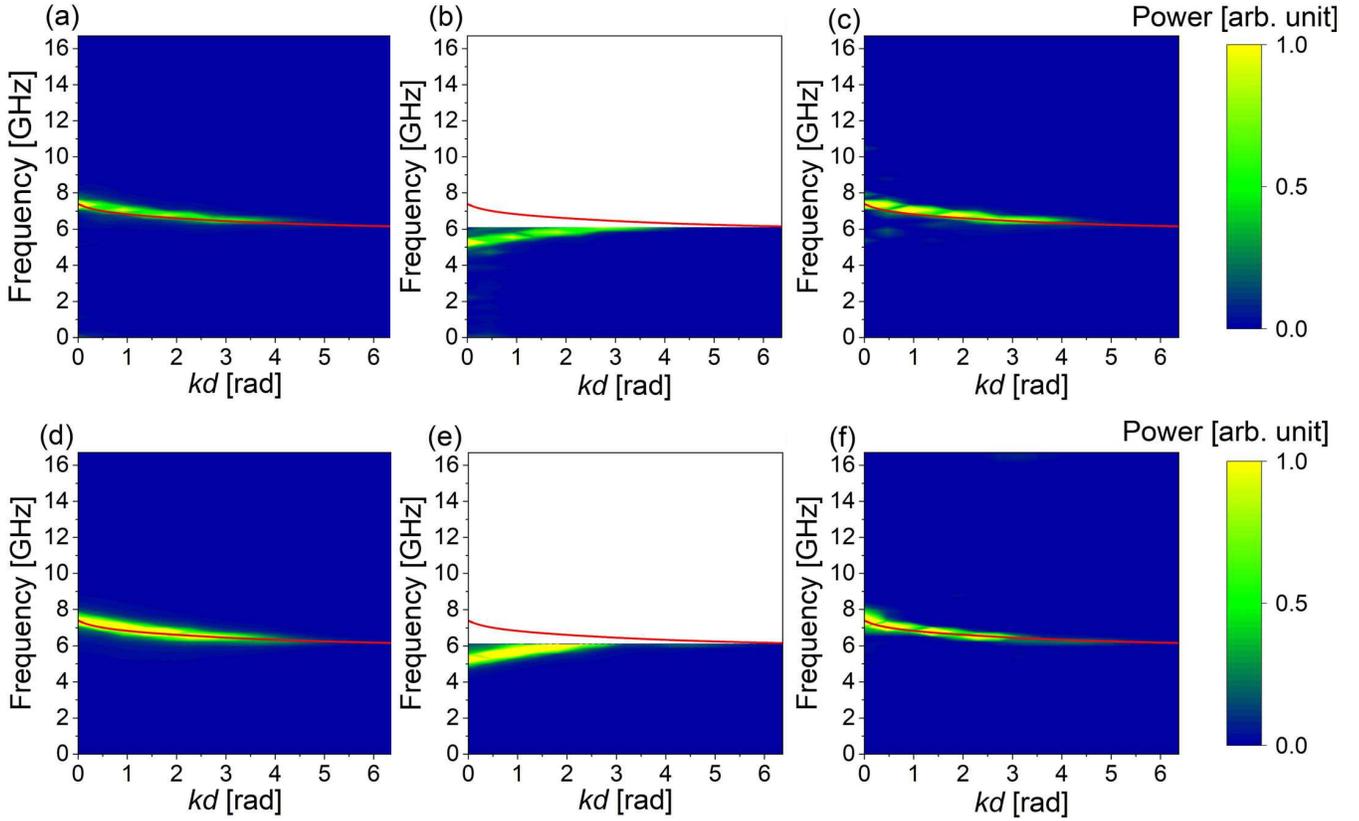}
    \caption{(a)--(c) Dispersion relations obtained from spatiotemporal waveforms shown in Figs. \ref{fig:STmap}(a)--(c).
    (d)--(f) dispersion relations obtained by analyzing the simulated waveforms corresponding to Figs. \ref{fig:STmap}(d)--(f).
    The red lines are the theoretical curves of the dispersion relation of the backward volume magnetostatic wave.
    The horizontal axis is the wavenumber multiplied by the thickness of the sample.}
    \label{fig:dispersion}
\end{figure*}

Figure \ref{fig:dispersion} shows the dispersion relations corresponding to the dataset shown in Fig \ref{fig:STmap}. 
Figure \ref{fig:dispersion}(a) was obtained from the data shown in Fig. \ref{fig:STmap}(a) by DFT in the time and space domain.
Moreover, Fig. \ref{fig:dispersion}(b) was obtained by DFT from the data shown in Fig. \ref{fig:STmap}(b).
The data in Fig. \ref{fig:dispersion}(c) were estimated via the LASSO method instead of DFT in the time domain.
Figure \ref{fig:dispersion}(b) shows that the information was only available up to
$6 \ \mathrm{GHz}$ due to the insufficient sampling rate of the data in Fig. \ref{fig:STmap}(b).
Therefore, signals appearing to exhibit the dispersion relation are folding noises bounded by the Nyquist frequency.
In Fig. \ref{fig:dispersion}(c), the same curve can be observed as in Fig. \ref{fig:dispersion}(a), which implies that sufficient 
information could be extracted from random sampling as shown in Fig. \ref{fig:STmap}(c).
Figures \ref{fig:dispersion}(d)--(f) show the dispersion relations calculated from the data shown in Figs. \ref{fig:STmap}(d)--(f),
the waveforms calculated by MuMax3.
The results of the simulation and the experiment were found to be in good agreement.

The excited modes were the backward volume magnetostatic waves,
which are mainly dominated by magnetic dipole interactions,
and they have a negative gradient of dispersion \cite{damon1961magnetostatic, hurben1995theory, hurben1996theory}.
The observed dispersion was in good agreement with the lowest order mode
shown in Fig. \ref{fig:dispersion} with red lines, and there were no peaks 
corresponding to the higher order modes.
This is because the higher order modes have nodes in the thickness direction, and
the Faraday rotation caused by the higher modes was mostly cancelled out through
the light transmittance.
The multiple branches seen in Figs. \ref{fig:STmap}(a) and (d) are due to the spin-wave echoes \cite{serga2010yig}.
\begin{figure}
    \centering
    \includegraphics[width=0.9\linewidth]{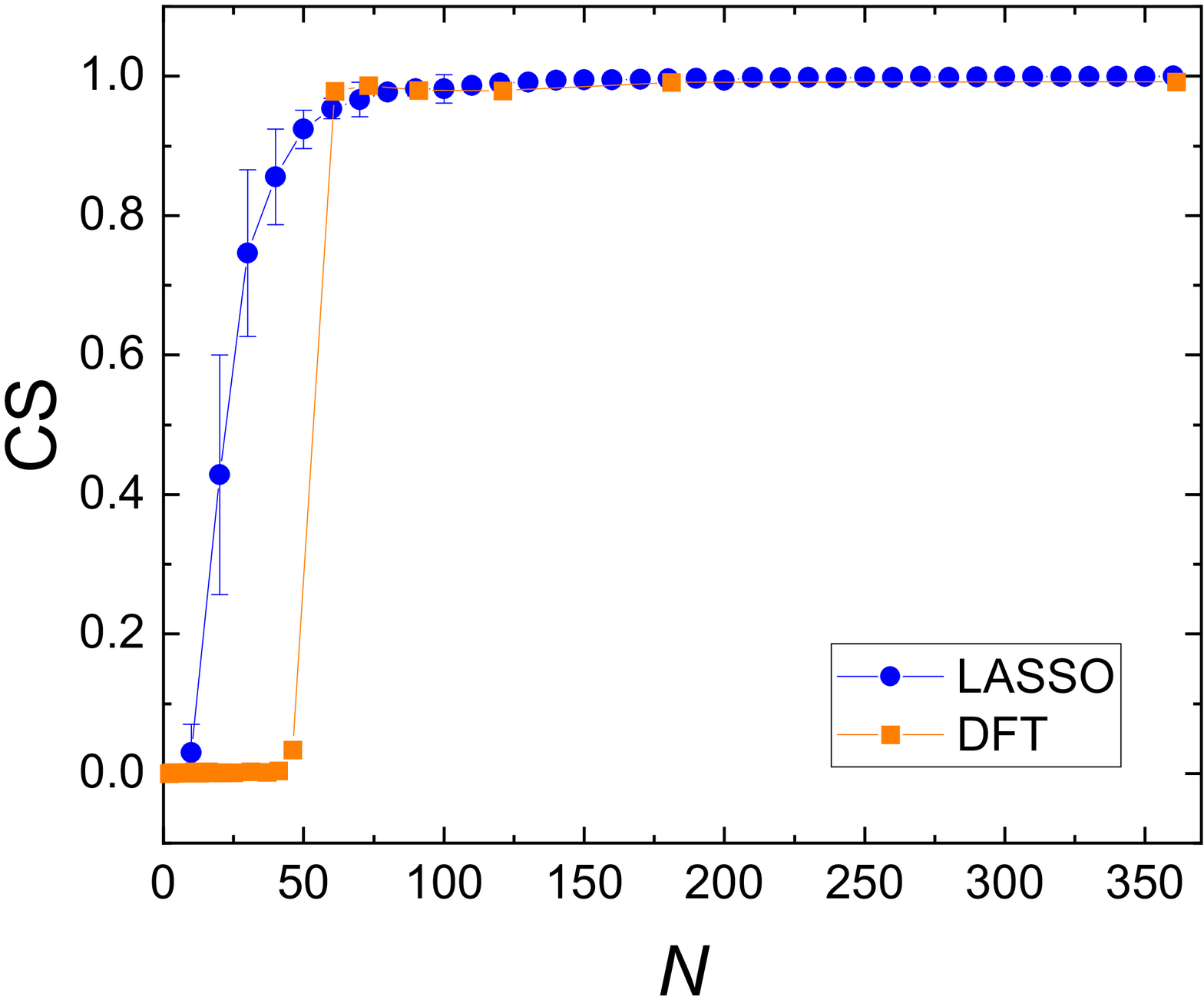}
    \caption{Cosine similarities of the dispersions at reduced $N$ to the dispersion at $N=361$. 
    Blue and orange dots represent the results of LASSO method and DFT, respectively. 
    The error bars of blue dots represent the standard deviations of 
    the CSs of 10 different sets of random sampling.}
    \label{fig:DFTvsLASSO}
\end{figure}

Figure \ref{fig:DFTvsLASSO} shows the 
$N$-dependence of the CSs between the dispersion obtained from the data with $N=361$ and the dispersions 
obtained from the data with reduced $N$.
For the DFT method, the points at $N \leq 46$ correspond to the conditions with $\Delta t \geq 0.08\ \mathrm{ns}$,
and their CSs were almost zero
because the Nyquist condition was not satisfied.
In contrast, the CSs of the results of the LASSO method were above $0.90$, 
even when $N$ was reduced to $50$.
Note that in the LASSO method, $N$
was reduced without decreasing the Nyquist frequency.

%% file: conclusion.tex
It is necessary to discuss in what systems LASSO can be applied.
Empirically, samplings 2--5 times the number of sparse 
coefficients is sufficient to reconstruct the spectrum using the $l_1$ norm \cite{lustig2007sparse}.
Based on this, even for a system with multiple modes, the number of required samples can be roughly estimated from the number of predicted peaks.
Furthermore, for a system with strong damping, the linewidth of the spectrum may be underestimated. 
This can be improved by setting the sampling time range to $T\approx 1/(\alpha f_0)$
where $f_0$ is the center frequency.

In conclusion, 
we demonstrated the fast acquisition of spin-wave dispersion by using compressed sensing.
Further, we quantitatively evaluated the effects of random sampling
on the results of the LASSO method.
This technique significantly reduced the measurement time 
for acquiring the dispersion relations.
Moreover, this method of applying compressed sensing to time-resolved pump-probe 
measurements is not limited to the 
magneto-optical imaging of spin waves, 
to various experiments based on pump-probe measurements,
such as observations via the electro-optical effects and the refractive index modulations.